\documentclass[sigconf,screen,review=false]{acmart}
\usepackage{graphicx}
\usepackage{wrapfig}
\usepackage{caption} 
\usepackage{multirow}
\usepackage{subfigure}
\usepackage{subcaption}
\usepackage{enumitem}
\usepackage{comment}
\usepackage{balance}
\usepackage{float}
\usepackage{wrapfig}

\AtBeginDocument{%
  \providecommand\BibTeX{{%
    \normalfont B\kern-0.5em{\scshape i\kern-0.25em b}\kern-0.8em\TeX}}}

\copyrightyear{2024}
\acmYear{2024}
\setcopyright{rightsretained}
\acmConference[CCEAI 2024]{2024 8th International Conference on Control Engineering and Artificial Intelligence}{January 26--28, 2024}{Shanghai, China}
\acmBooktitle{2024 8th International Conference on Control Engineering and Artificial Intelligence (CCEAI 2024), January 26--28, 2024, Shanghai, China}\acmDOI{10.1145/3640824.3640840}
\acmISBN{979-8-4007-0797-1/24/01}

\usepackage{makecell}

\usepackage{algorithm}
\usepackage[noend]{algpseudocode}

\usepackage{xcolor}

\newcommand{\Title}{\emph{Decaffe}: DHT Tree-Based Online Federated Fake News Detection}

\usepackage{acro}

\acsetup{format/first-long =\itshape} 

\DeclareAcronym{FL}{
short = FO,
long = federated optimization
}
\DeclareAcronym{PFL}{
short = PFO,
long = personalized federated optimization
}
\DeclareAcronym{MSN}{
short = MSN,
long = mobile social network,
short-plural-form = MSNs,
long-plural-form = mobile social networks
}
\DeclareAcronym{PS}{
short = PS,
short-plural-form = PSs,
long = parameter server,
long-plural-form = parameter servers
}
\DeclareAcronym{non-IID}{
short=non-IID,
long=non-independent-and-identically-distributed
}
\DeclareAcronym{DHT}{
short = DHT,
long = distributed hash table
}
\DeclareAcronym{RL}{
short=RL, long=reinforcement-learning
}
\DeclareAcronym{DDQL}{
short=DDQL,
long=Double Deep Q-Learning
}
\DeclareAcronym{DNN}{
short=DNN,
short-plural-form=DNNs,
long=deep neural network
}
\DeclareAcronym{CoPT}{
short= CoPT, 
long= copies of parameters transmitted
}
\DeclareAcronym{IFCA}{
short = IFCA,
long = Iterative Federated Clustering Algorithm
}
\DeclareAcronym{CFL}{
short = CFL,
long = Cluster-based Federated Learning
}
\DeclareAcronym{FedAvg}{
short = FedAvg,
long = Federated Averaging
}
\DeclareAcronym{CS}{
short = CS,
long = cosine similarity
}
\DeclareAcronym{DoSD}{
short = DoSD,
long = density of selected devices
}
\DeclareAcronym{DACN}{
short = DACN,
long = Double Actor-Critic Networks
}
\DeclareAcronym{ACN}{
short = ACN,
long = Actor-Critic Networks
}
\DeclareAcronym{DDPG}{
short = DDPG,
long = Deep Deterministic Policy Gradient
}
\DeclareAcronym{IRB}{
short = IRB,
long = inverted residual block,
long-plural-form = inverted residual blocks }
\DeclareAcronym{FCL}{
short = FCL,
long = fully connected layer,
short-plural-form = FCLs,
long-plural-form = fully connected layers
}
\DeclareAcronym{LSTM}{
short = LSTM,
long = Long Short-term Memory
}
\DeclareAcronym{DRL}{
short = DRL,
long = deep-reinforcement-learning
}
\DeclareAcronym{NLP}{
short = NLP,
long = natural language processing}
\DeclareAcronym{BERT}{
short = BERT,
long = Bidirectional Encoder Representations from Transformers
}
\DeclareAcronym{ML}{
short = ML,
long = machine learning
}
\DeclareAcronym{PCA}{
short = PCA,
long = principal component analysis
}
\DeclareAcronym{MmFLA}{
short = MmFLA,
long = Multi-model Federated Learning Algorithm
}
\DeclareAcronym{GMI}{
short = GMI,
long = global model identity,
short-plural-form = GMIs,
long-plural-form = global model identities
}
\DeclareAcronym{CCA}{
short = CCA,
long = converged classification accuracy
}

\newcommand{\systemName}{DHT Tree-Based Online Federated Fake News Detection}
\newcommand{\systemNameAbbr}{\emph{Decaffe}}

\newcommand{\CSB}{the centralized server-based model fine-tuning}
\newcommand{\DSL}{the decentralized serverless model fine-tuning}

\usepackage{titlesec}
\titlespacing*{\section}
{0pt}{0.2ex}{0.1ex}
\titlespacing*{\subsection}
{0pt}{0.1ex}{0.1ex}
\titlespacing*{\subsubsection}
{0pt}{0.05ex}{0.05ex}
\titlespacing*{\paragraph}
{0pt}{0.05ex}{0.05ex}
\titlespacing*{\enumerate}
{0pt}{0.05ex}{0.05ex}

\begin{document}

\title{\Title}

\author{Cheng-Wei Ching}
\orcid{0000-0001-6621-4907}
\affiliation{%
  \institution{University of California Santa Cruz}
  \streetaddress{Santa Cruz 1156 High Street}
  \city{Santa Cruz, CA 95064}
  \country{USA}}
\email{cching1@ucsc.edu}

\author{Liting Hu}
\orcid{0009-0007-7222-5507}
\affiliation{%
  \institution{University of California Santa Cruz}
  \streetaddress{Santa Cruz 1156 High Street}
  \city{Santa Cruz, CA 95064}
  \country{USA}}
\email{liting@ucsc.edu}

\renewcommand{\shortauthors}{Ching, et al.}

\begin{abstract}

The proliferation of mobile social networks (MSNs) has transformed information dissemination, leading to increased reliance on these platforms for news consumption. However, this shift has been accompanied by the widespread propagation of fake news, posing significant challenges in terms of public panic, political influence, and the obscuring of truth. Traditional data processing pipelines for fake news detection in MSNs suffer from lengthy response times and poor scalability, failing to address the unique characteristics of news in MSNs, such as prompt propagation, large-scale quantity, and rapid evolution. This paper introduces a novel system named {\systemNameAbbr} – a DHT Tree-Based Online Federated Fake News Detection system. {\systemNameAbbr} leverages distributed hash table (DHT)-based aggregation trees for scalability and real-time detection, and it employs two model fine-tuning methods for adapting to mobile network dynamics. The system's structure includes a root, branches, and leaves for effective dissemination of a pre-trained model and ensemble-based aggregation of predictive results. {\systemNameAbbr} uniquely combines centralized server-based and decentralized serverless model fine-tuning approaches with personalized model fine-tuning, addressing the challenges of real-time detection, scalability, and adaptability in the dynamic environment of MSNs.

\end{abstract}



\maketitle

\section{Introduction}
\label{sec: introduction}

\noindent With the boom of the internet and the rapid popularity of smartphones, \acp{MSN} have become an important platform for information dissemination. More and more people tend to seek out and consume news directly from social media in \acp{MSN} rather than the traditional news media. 
This is because 
(i) it is easier to consume news on social media than the traditional news media such as newspapers or television news; 
and (ii) users like to comment on, share, and discuss the news with friends or other users in \acp{MSN}. 
However, such high openness and autonomy of \acp{MSN} lead the widespread of fake news.
For instance, the government banned several What's App groups for spreading fake news about distorting the information of Agnipath Scheme For instance, a survey conducted by Statista in 2022 indicates that two-third of US adults say they have come across false information on social media \cite{readfakenews}; and \acp{MSN} are considered the least trusted news source worldwide in 2022 \cite{leasttrust}.


Fake news spread in \acp{MSN} poses a serious negative impact on individuals and society.
First, fakes news is found easy to cause public panic, which results in mass panic behaviors. 
Second, fake news hold much sway over the trend of political events. Propagandists take advantage of the fact that people consider a lie truthful when the lie is repeated enough times over social media. Many studies show that during the significant political events, such as the US Presidential elections and the Brexit referendum, user interactions with false content rose
steadily on well-known social media \cite{di2021fake,allcott2019trends,bastos2019brexit}.
Third, fake news makes it harder for people to see the truth. It triggers people's distrust and makes them confused, impeding their abilities to differentiate the truth from the falsity.

To help mitigate the negative effects caused by fake news, there is an urgent need to quickly detect fake news circulated in \acp{MSN} early in its propagation before it reaches a broad audience. There are three critical characteristics that differ news in \acp{MSN} from the traditional news media. First, \textit{the prompt propagation}. When news is posted in \acp{MSN}, it spreads to numerous people in a very short period of time and will soon reach its peak rate of comments, retweets or share \cite{wu2018tracing}. Second, \textit{the large-scale quantity of news}. People nowadays spend a large amount of time browsing, sharing, and comments on news in \acp{MSN}, which creates a large-scale quantity of data in a short period of time \cite{huang2017time}. Lastly, \textit{the rapid evolution of news}. News in \acp{MSN} is usually forwarded or commented by people, and then rapidly converted into different context containing other details or personal opinions
\cite{di2021fake}.

Unfortunately, the traditional data processing pipeline for fake news detection systems \cite{agrawal2018rheem,hung2018wide} in \acp{MSN} mainly follow the classic big data analytics pipeline. Geo-distributed web servers first collect a large amount of users' microblog data and then transmit the data to a centralized storage tier (e.g., MongoDB). Upon the arrival of all the microblog data, big data analytics engines (e.g., Spark Streaming with MLlib \cite{SparkML}) begin processing and analyzing, and finally report the authenticity of the data.
The pipeline above has two obvious disadvantages.
One is the lengthy response time.
A large amount of time is necessary to transmit microblog data from geo-distributed web servers to a central storage tier for further processing and analyzing, ending up returning the results.
Such a disadvantage cannot tackle the prompt propagation of news in \acp{MSN} since the results may have expired.
Another is the inferior scalability. The centralized data processing is ill-scalable to large-scale data, which fails to handle the large-scale quantity of news in \acp{MSN}. 
Two disadvantages above are the major motivation of this paper.

\begin{figure}
    \centering
    \includegraphics[width=5.5cm]{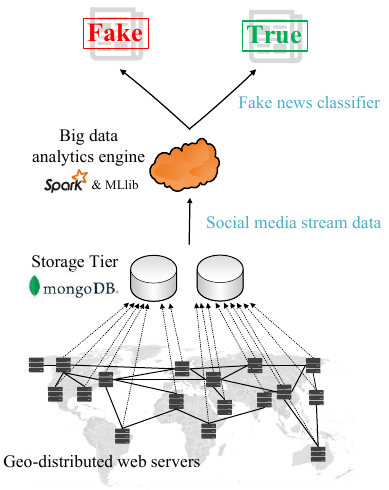}
    \caption{Traditional data processing pipeline.}
    \label{fig: existing system}
\end{figure}

To this end, we propose a novel {\systemName} (\systemNameAbbr). The key features of {\systemNameAbbr} are twofold. On the structural side, {\systemNameAbbr} establishes \ac{DHT}-based aggregation trees to realize scalability and real-time fake news detection in \acp{MSN}. On the model fine-tuning side, {\systemNameAbbr} devises two model fine-tuning methods, \textit{\CSB} and \textit{\DSL}, to fine tune \ac{ML} model \ac{BERT} for fake news detection and adapt to the mobile network dynamics. 



The novelty of {\systemNameAbbr} is threefold.
First, {\systemNameAbbr} uses \ac{DHT}-based aggregation tree to realize {\CSB}, {\DSL}, and ensemble-based model inference. 
Then, a simple yet effective differentiation method is designed to determine an appropriate model fine-tuning method.
Lastly, {\systemNameAbbr} incorporates personalized model fine-tuning to take into account the heterogeneous data and detection requirements among different mobile devices.


\section{Background and Preliminaries}
\label{sec: preliminaries}




\subsection{Traditional data analytics pipeline}

As shown in Fig. \ref{fig: existing system}, existing studies \cite{oh2021network,jonathan2018multi,hu2019spear} mostly reply on a centralized data analytics pipeline for fake news detection in  \acp{MSN}. 
Specifically, geo-distributed web servers collects microblog data from the users or edge servers and upload the microblog data collected to a storage tier (e.g., MongoDB). Then, the big data analytics engines (e.g., Spark \cite{SparkML} with MLlin) process the microblog data and return the authenticity of the microblog data.

Such a centralized architecture may run well in tackling non-time-sensitive applications. However, when it comes to the fake news detection in \acp{MSN}, fake news evolves fast and usually has distinct topics and are posted and disseminated real-time, which makes the predictive results from the centralized architecture obsolete and inaccurate. This is because of (1) \textbf{high response time}. Not until the microblog data is uploaded do data analytics engines begin processing and analyzing, which cannot meet the requirement of real-time response; and (2) \textbf{Poor-scalability}. 

If a new kind of topics of news emerges, the centralized architecture takes a large amount of time to adapt to it, which cannot offer accurate response when fake news evolves or changes its context.

\subsection{Preliminaries}

\noindent\textit{Decentralized Overlay in Pastry.}
\label{subsec: dht}
Suppose there are $N$ nodes participating in Pastry. 
On the initialization of Pastry, each node is assigned with an individual identifier (i.e., \texttt{nodeId}). The \texttt{nodeId} is used for node identification and message route. Given a message and a key, the message is guaranteed to reach the node with a \texttt{nodeId} numerically closet to the given key at most $O(\lceil\log_{2^b} N \rceil)$, where $b = 4$ by default. Moreover, each node maintains two data structures, \textit{a routing table} and \textit{a leaf set}, in support of message routing, self-organization, and fault recovery functionalities.

We first go over the designs of routing tables and leaf sets. Routing tables in each node in Pastry are composed of node prefixes, such as IP address, latency, Pastry \texttt{nodeId}, arranged in rows by the common prefix length. On the other hand, leaf sets in each node in Pastry consists of a fixed number of nodes, 24 by default, whose \texttt{nodeIds} are numerically closet to the node. Leaf sets help nodes rebuild routing tables when nodes fail.
Both routing tables and leaf sets are critical to the message routing in Pastry.


\begin{figure}
    \centering
    \includegraphics[width=5.5cm]{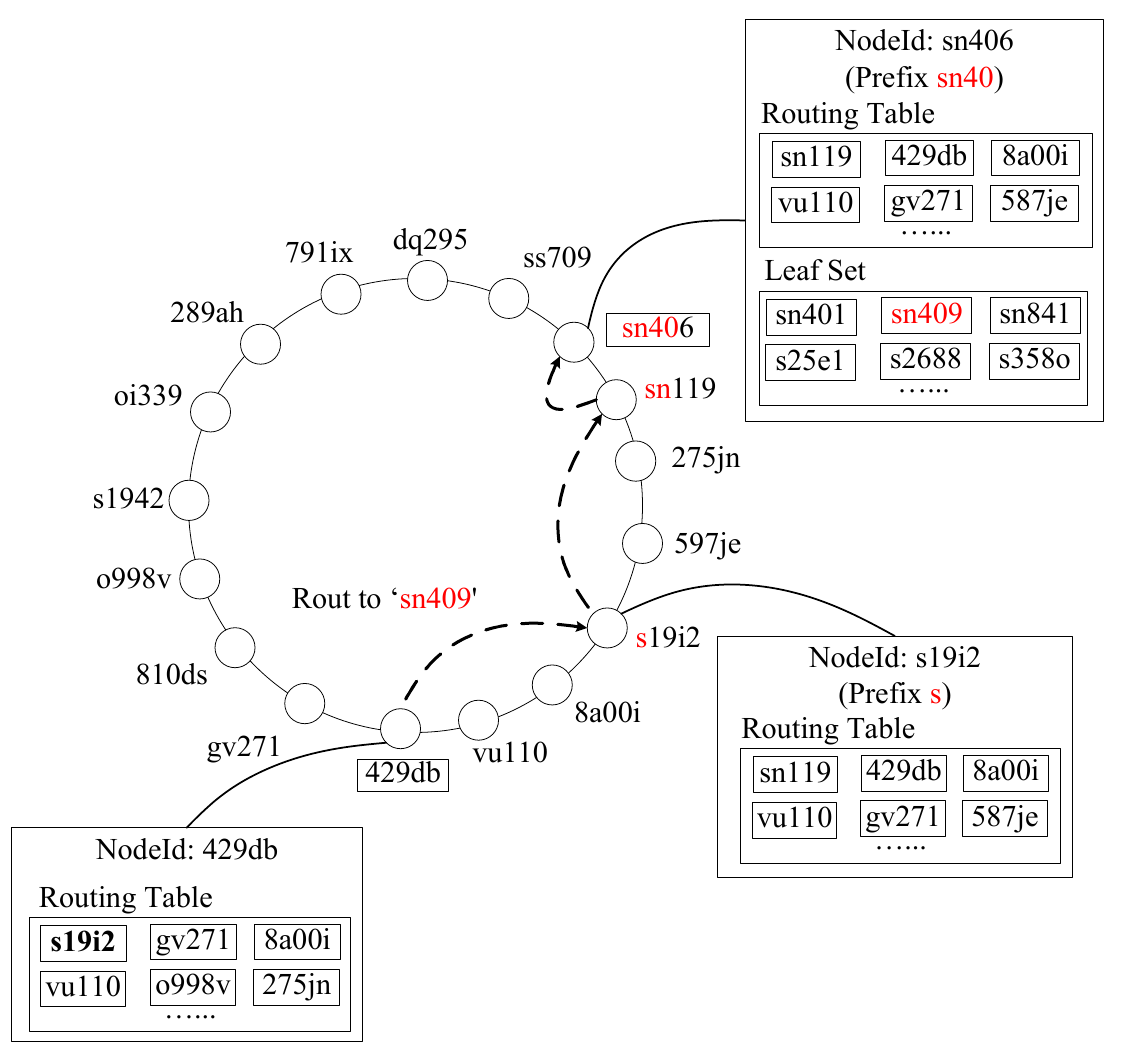}
        \caption{Message routing in Pastry.}
    \label{fig: message routing in pastry}

\end{figure}

For the messaging routing in Pastry, messages are routed in a greedy fashion. A simple example is depicted in Figure \ref{fig: message routing in pastry}. 
Suppose a message with a key of \texttt{sn409} is generated in the node \texttt{429db}, the node \texttt{429db} forwards the message to a node in its routing table whose common prefix shares at least one digit with the given key, where it is the \texttt{s19i2} in this example.

If such a node does not exist, the node \texttt{429db} forwards the message to a node whose common prefix 
shares with the given key at least as long as the local node, and is numerically closer to the given key than the local \texttt{nodeId}. 
Then, the node \texttt{s19i2} follows the same logic to forward the message. The message routing ends until the node whose \texttt{NodeId} is numerically closest to the key receives the message. In this example, the node \texttt{sn409} receives the message and the message routing is complete.

\noindent\textit{Group Management by Scribe.}
\label{subsec: scribe}
Scribe is a communication system built upon Pastry for the management of application-level groups \cite{castro2002scribe}. 
Each application-level group in Scribe is built upon a logical spanning tree that consists of multiple members (i.e., nodes in Pastry). When nodes leave or rejoin the overlay, Scribe manages group sizes accordingly, and supports rapid switch for the management of group membership \cite{rowstron2001pastry, castro2002scribe}. For the group generation in Scribe, Scribe inherits the pseudorandom Pastry key to name a group, called \texttt{groupId}, where the \texttt{groupId} is usually the hash of the group's textual name concatenated with its creator's name. In practice, a Scribe node uses the messaging routing in Pastry to route a \texttt{CREATE} message with the \texttt{groupId} as the key. The node whose \texttt{nodeId} is the numerically closet to the key becomes the root of the spanning tree for the application-level group. 

\section{{\systemNameAbbr} Design}
\label{sec: algorithm description}


In this section, we introduce the {\systemNameAbbr} system, discuss each
two phases in {\systemNameAbbr}, and outline the details
of workflows in the {\systemNameAbbr} system.

\subsection{Overview}
\label{subsec: overview}
{\systemNameAbbr} operates in two phases. The first phase is the construction of \ac{DHT}-based aggregation trees, and the second phase is the fake news detection application and online fine-tuning. 
In the first phase, {\systemNameAbbr} constructs \ac{DHT}-based aggregation trees, and then uses the constructed trees to execute online model fine-tuning and ensemble-based inference in the second phase.

Each \ac{DHT}-based aggregation tree consists of a root, multiple branches, and multiple leaf nodes. 
The root serves as the main control flows for the entire aggregation tree, such as the dissemination of initial model and intermediate fine-tuning model to the leaf nodes and the aggregation of final predictive results or intermediate fine-tuning models from the leaf nodes. 
The branches are responsible for two main works. One is the dissemination as what the root is supposed to do. Another is the intermediate aggregation of intermediate fine-tune models and predictive results such that the network bandwidth can be further conserved. 
The leaf nodes are responsible for performing local model fine-tuning for fake news detection. When new data comes to the leaf nodes, they first infer the new data with local fine-tuned model. 

In the second phase, the leaf nodes execute two tasks concurrently. One task is the online mode fine-tuning and the other task is the ensemble-based model inference. The online model fine-tuning is launched when the current models in the leaf nodes are outdated. 
The ensemble-based model inference is for inferring the new coming data on the leaf nodes and the predictive results will be aggregated along the aggregation tree to the root. The final predictive results are attained in an ensemble manner.

\subsection{Construction of \ac{DHT}-based aggregation trees}

\noindent\textit{Root.}
The root serves as the main control flows for the entire aggregation tree, including 
1) the dissemination of initial pre-trained model and intermediate fine-tuning model to the leaf nodes, and 2) the aggregation of final predictive results or intermediate fine-tuning models from the leaf nodes. 

The root uses a DHT-based aggregation tree as the
main approach to communicate with the leaf nodes. The DHT-based aggregation trees are constructed as follows:
\begin{enumerate}[start = 1, align = left, leftmargin = *, wide = 2mm, label = Step \arabic*:, ref= \arabic*, topsep=0pt]
    \item The first step is the construction of a peer-to-peer overlay network leveraging Pastry \cite{rowstron2001pastry}. Each node is first assigned a unique \texttt{nodeId} in a circular \texttt{nodeId} space with the range from $0$ to $2^{128} -1 $, where the \texttt{nodeId} of each node is uniformly distributed. Given a message and a key in a node, the message can be guaranteed to be routed to the node whose \texttt{nodeId} is numerically closest to the given key at most $O(\log_{2^b}N)$ hops, where $b=4$ by default.
    \item The second step builds a multicast tree leveraging Scribe \cite{castro2002scribe}. Each node in the overlay is able to generate a group with a \texttt{groupId} that is usually the groups's textual name concatenated with its creator's name. Upon the generation of a group, each node can join the group by routing a \texttt{JOIN} message towards the \texttt{groupId}. Multicast messages can be sent from the root to any member node at most $O(\log N)$ hops.
    \item The root works with the branches on the aggregation of online model fine-tuning and ensemble-based model inference. For online model fine-tuning, the root and the branches aggregate the intermediate fine-tuned model, and the root disseminates the finally aggregated fine-tuned model back to the leaf node. 
    For ensemble-based model, the root and the branches aggregate the predictive results from the leaf nodes, and the root provides the end users with final predictive results using the ensemble-based model inference.
\end{enumerate}

\noindent\textit{Leaf nodes.} 
The leaf nodes execute two major tasks concurrently: the online model fine-tuning and fake news detection. The root will disseminate the initial pre-trained model to the leaf nodes. Upon receiving the initial model, the leaf nodes perform local model fine-tuning with local data. When new data coming from the users, the leaf nodes infer the new data with the local fine-tuned model and send the predictive results to the root along the aggregation tree. Once the local fine-tuned model is outdated, the leaf nodes launch online model fine-tuning to update the local fine-tuned model. 

\noindent\textit{Self-adjustable aggregation trees.}
The aggregation trees in {\systemNameAbbr} can self-tune the tree structure level by manipulating the fanout. Specifically, the fanout is the maximum degree of each node in the aggregation tree. For example, when the fanout is 2, then the aggregation tree is as much as a binary tree. When the fanout is $N$, where $N$ is the number of nodes, then the aggregation tree changes to a star graph. The value of fanout can be adaptive to the characteristics of applications. Suppose an application is latency-intensive. The fanout of aggregation trees can be as small as possible such that the height of aggregation trees is small, which take less time on disseminating and aggregating information. On the other hand, higher aggregation trees are more fault-tolerant. Each node in {\systemNameAbbr} is constantly receiving a existence message from its parent. Then nodes will route the \texttt{JOIN} message again to rejoin the aggregation tree when their parents fail to transmit existence messages. When a aggregation tree has a smaller fan-out
and a larger height, the failure of one node can only affect a small number of nodes.

\subsection{Online model fine-tuning and ensemble-based model inference}

\noindent\textit{Online model fine-tuning.}
Inspired by the current success of \ac{BERT} on \ac{NLP}, 
we use \ac{BERT} as the base model. BERT is a language representation model designed to pre-train deep bidirectional representations by jointly conditioning on both left and right context in all layers \cite{kenton2019bert}.
In practice, we add one more output layer in the pre-trained \ac{BERT} and the leaf nodes use local data to fine-tine the pre-trained \ac{BERT}. The pre-trained \ac{BERT} model is fine-tuned as follows:
\begin{enumerate}[start = 1, align = left, leftmargin = *, wide = 2mm, label = Step \arabic*:, ref= \arabic*, topsep=0pt]
    \item When receiving new data from the users, the leaf nodes first use the standard WordPiece embeddings to tokenize the new data into multiple tokens. 
    \item Feed the tokens to the pre-trained \ac{BERT} and obtain the predictive results.
    \item Calculate the loss between the ground truth and the predictive results and the corresponding gradients.
    \item Upload the intermediate fine-tuned models to the root or to the next leaf nodes along the aggregation tree, depending on the selected model fine-tuning rules, {\CSB} or {\DSL}. 
\end{enumerate}

\begin{figure}
    \centering
    \includegraphics[width=0.7\linewidth]{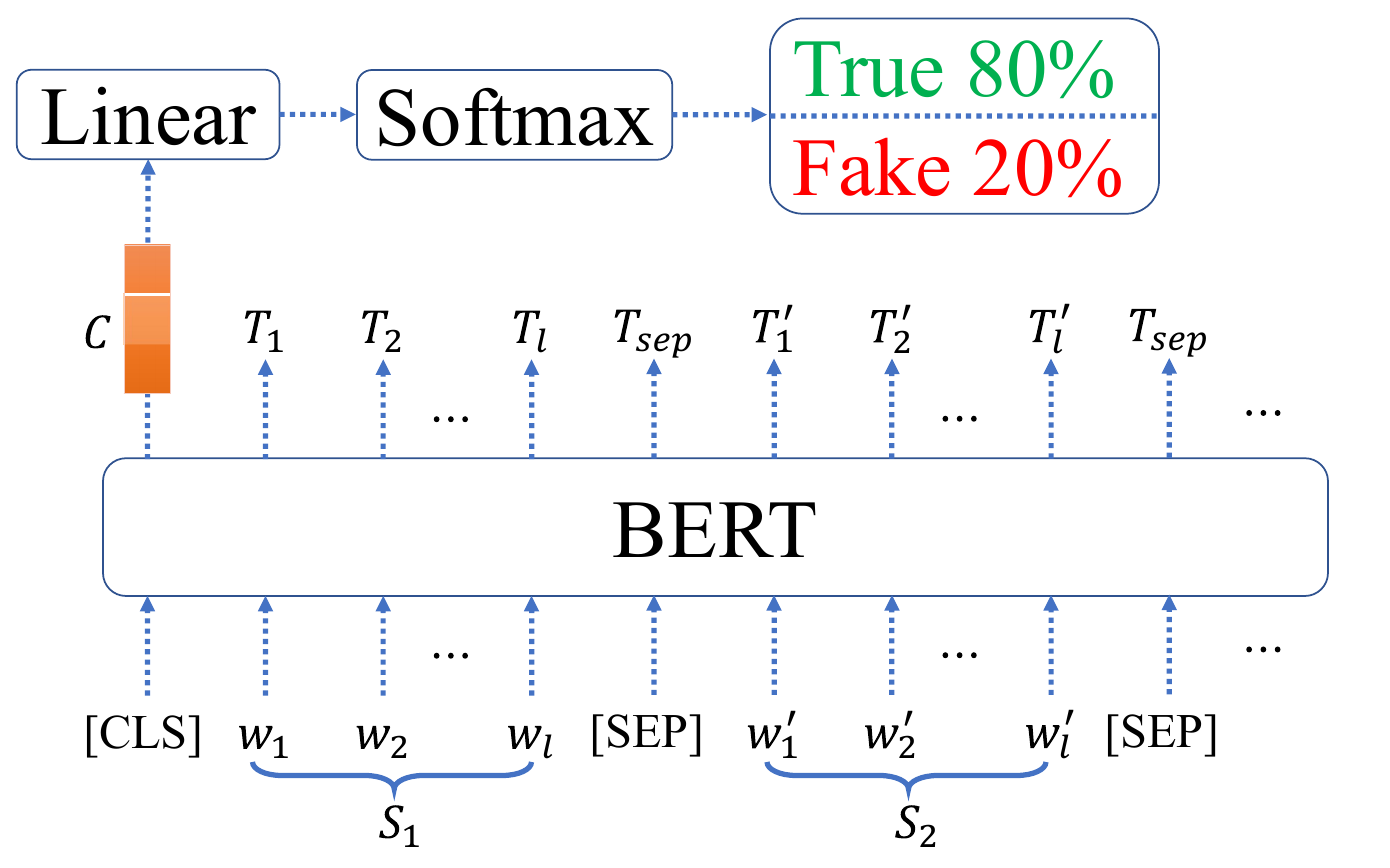}
    \caption{The \ac{BERT} model structure}
    \label{fig: bert model}
\end{figure}


\noindent\textit{Model structure and loss function.}
The \ac{BERT} model is shown in Fig. \ref{fig: bert model}. In particular, the input of \ac{BERT} includes \texttt{CLS} the classification token, \texttt{SEP} the separation token, $w_1, \cdots, w_l$ and $w'_1, \cdots, w'_l$ that represents the word in sentence $S_1$ and $S'_2$, respectively. The output of \ac{BERT} consists of $T_1, \cdots T_l$ and $T'_1, \cdots T'_l$ the hidden vector for input $w_1, \cdots, w_l$ and $w'_1, \cdots, w'_l$, respectively, and $C$ and $T_{sep}$ represent the hidden vector of the input \texttt{[CLS]} and \texttt{[SEP]}.

We use the hidden vector $C$ as sequence representation for fake news detection. The major parameters for federated-based fine-tuning is the weights in \texttt{Linear} $w_{cla} \in \mathbb{R}^{L\times H}$, where $L\in \{0,1\}$ is the number of labels and in fake news detection, $H$, the dimension of the hidden vector $C$.

For the loss function, we leverage the loss function in \ac{PFL}~\cite{ching2023dual,ching2020optimal} in order to retain the data characteristics in the different leaf nodes and use federated-based method to perform online model fine-tuning. Therefore, the loss function for the leaf node $n$ to perform online model fine-tuning is as follows:
\begin{align}
    l_n(w_{glo}) &\coloneqq \mathbb{E}_{(x,y)\sim\mathcal{D}_n}\big[\sum_{i\in \{0,1\}}\textbf{1}_{y=i}\log f_i(x;w_{cla})\big] + \frac{\lambda}{2}\|w^n_{per} - w_{cla}\|,
    \label{eq: loss for bert in leaf node}
\end{align}
where $(x,y) \sim \mathcal{D}_n$ denotes the data pair in local data $\mathcal{D}_n$, $w^n_{per}$, the personalized fine-tuned model in the leaf node $n$, $w_{glo}$, the global fine-tuned model, $f_i(x;w_{cla})$, the loss function for class $i$ parameterized by $w_{cla}$ to the sample $x$.
The leaf nodes will follow the above loss function to obtain intermediate fine-tuned models. 

\noindent\textit{Centralized server-based model fine-tuning.}
We see that the leaf nodes only optimize with respect to their local data. However, due to the characteristic of the rapid evolution of news, applying only one fine-tuned model for detection is insufficient. Therefore, the leaf nodes upload the intermediate fine-tuned models with respect to their loss to the root along the aggregation tree so that the leaf nodes can obtain a fine-tuned model that have better generalized performance for varying new data from \acp{MSN}.

As shown in Fig. \ref{fig: existing system}, {\CSB} relies on the communication between the root and the leaf nodes. Specifically, in communication round $t$, each leaf node $n\in N$ use the local new data $\mathcal{D}_n$ to perform local model fine-tuning and obtain the intermediate fine-tuned models $w^{n,t}_{cla}$. Then, the intermediate fine-tuned models are uploaded to the parent nodes of the corresponding leaf nodes. The parent nodes perform intermediate aggregation as follows:
$w^{n,t}_{cla} = \sum_{j\in \mathcal{C}_n} \frac{w^{j,t}_{cla}}{|\mathcal{C}_n|}$,
where $\mathcal{C}_n$ represents the child nodes of the node $n$, and $|\cdot|$, the cardinality operator. The intermediate aggregation can reduce the communication overhead by a factor of $O(\log N)$ since 
it maintains at most $O(\log N)$ instead of $N$ point-to-point connections for $N$ leaf nodes.

When receiving the intermediate fine-tuned models from its child nodes, the root $r$ performs final model fine-tuning to yield a new global fine-tuned model as follows:
$w^{t+1}_{cla} = w^t_{cla} - \eta \sum_{j\in \mathcal{C}_r} \frac{w^{j,t}_{cla}}{|\mathcal{C}_r|}.$
Lastly, the root uses the aggregation tree to multicast the updated fine-tuned model to each leaf node. Once the leaf nodes require to perform online model fine-tuning again, the process above is repeated again.

We can observe that an one-trip communication from leaf nodes to the root and back to leaf nodes is necessary for {\CSB}. In spite of the simplicity of {\CSB}, the entire aggregation tree easily suffer from communication and computational bottlenecks that happen to each member node~\cite{kuo2021energy}, especially in a mobile social network, where the large-scale quantity of news and the rapid quantity of data are two critical characteristics of \acp{MSN}. 
Moreover, to secure the user data from the model inversion attacks that reproduce the data for model fine-tuning \cite{ching2020model,ching2021icccn,totoro}, some security-intensive applications incorporate random noises into model fine-tuning \cite{abadi2016deep,totoro}. Such a random noise is proved to pose negative impact on predictive performance.


\noindent\textit{The decentralized serverless model fine-tuning.}
To mitigate the possible communication and computational bottleneck, we device {\DSL}. Specifically, the leaf nodes first perform local model fine-tuning and then upload their gradients to one of leaf nodes that \textit{have social links one another}~\cite{ching2020energy, ching2021efficient}. After receiving the intermediate fine-tuned models from their trust leaf nodes, the leaf nodes aggregate the intermediate fine-tuned models and continue uploading gradients until the root receives the intermediate fine-tuned models.

As shown in Fig. \ref{fig: message routing in pastry}, {\DSL} relies on the peer-to-peer communication between leaf nodes and ends in the root. Specifically, in communication round $t$, each leaf node $n\in N$ use the local new data $\mathcal{D}_n$ to perform local model fine-tuning and obtain the intermediate fine-tuned models $w^{n}_{cla}$. Then, the leaf nodes send the intermediate fine-tuned models to the friend leaf nodes that has social links with them. The friend leaf nodes perform intermediate aggregation as follows:
$w^{i,t}_{cla} = \sum_{j\in \mathcal{F}_i} \frac{w^{j,t}_{cla}}{|\mathcal{F}_i|},$
where $\mathcal{F}_i$ represents the leaf nodes that have social links with the leaf node $i$.

\begin{figure*}[t]
    \centering
    \subfigure[COVID19 News]{\label{subfig: covid acc}\includegraphics[width=.2\linewidth]{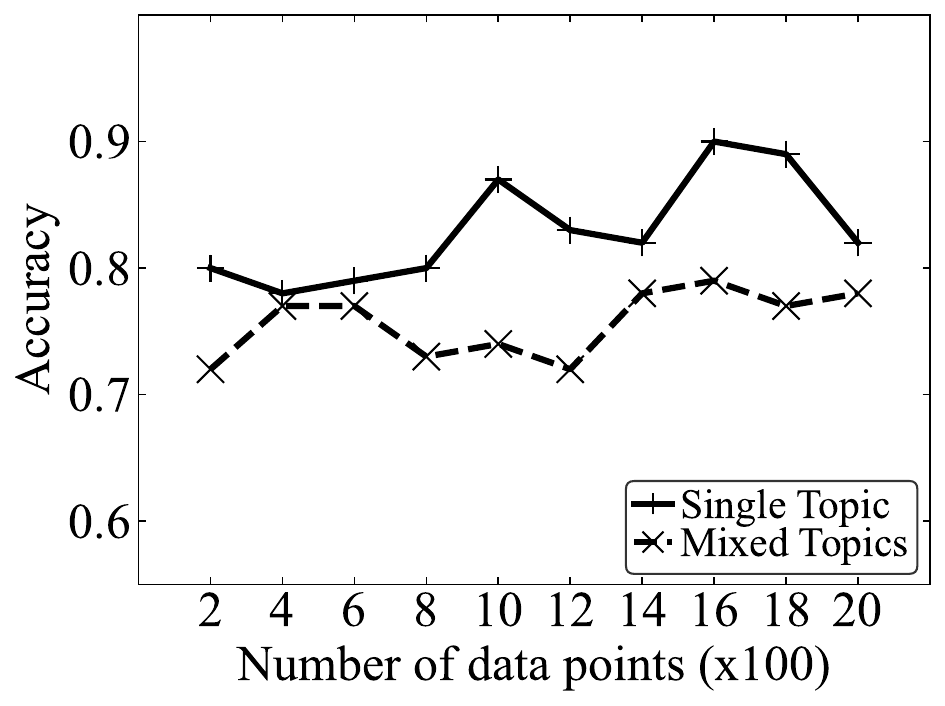}}
    \subfigure[Vaccination News]{\label{subfig: vaccination acc}\includegraphics[width=.2\linewidth]{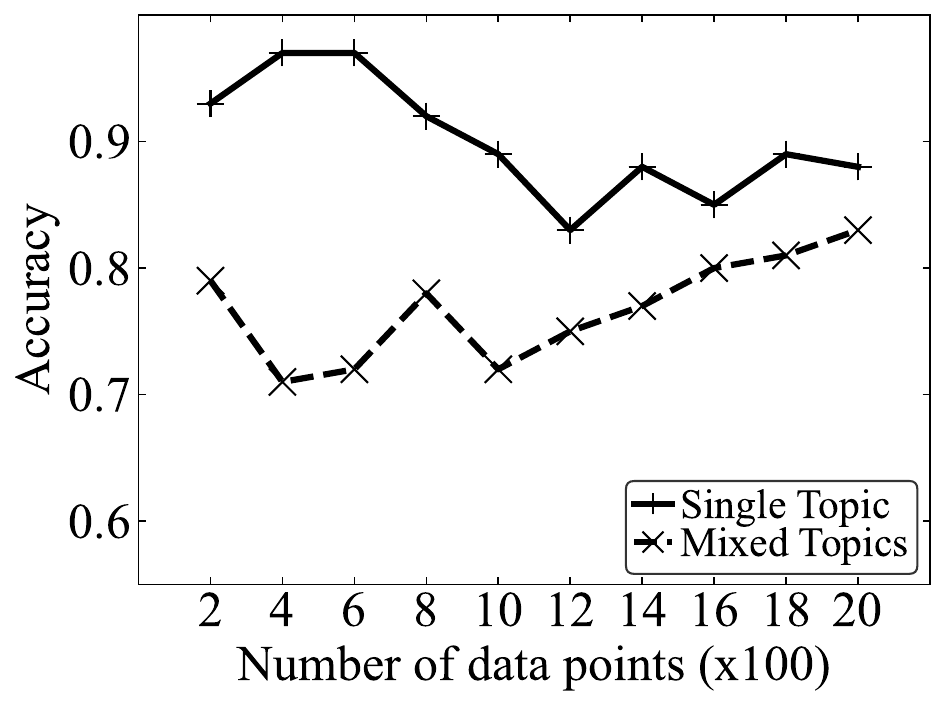}}
    \subfigure[Election News]{\label{subfig: election acc}\includegraphics[width=.2\linewidth]{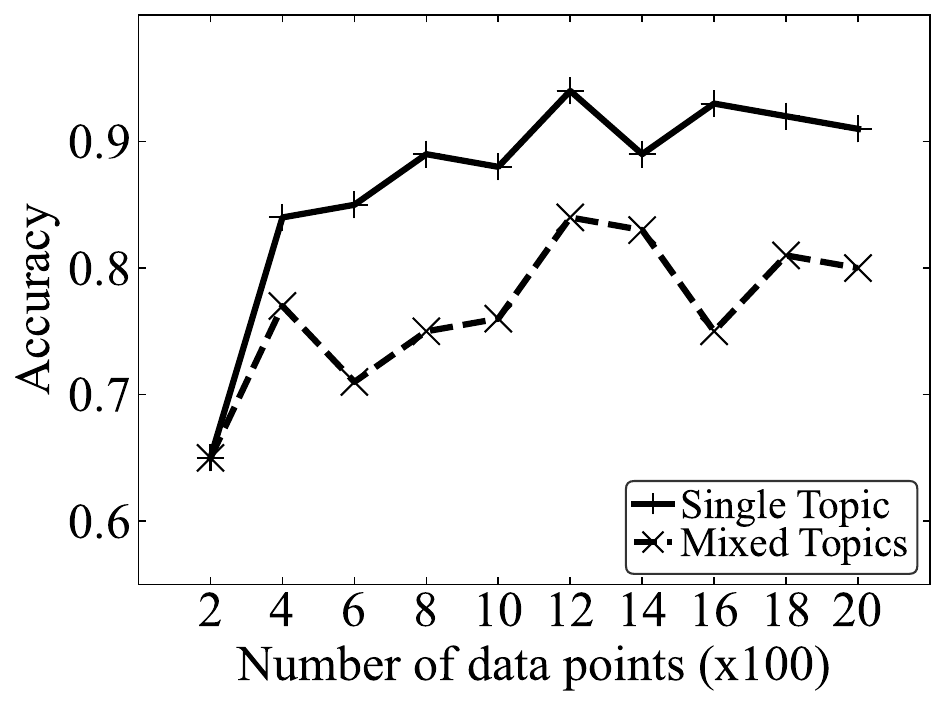}}
    \caption{Accuracy on three news datasets: COVID19 news, vaccination news, and election news.}
    \label{fig: three news dataset on acc}
\end{figure*}

\begin{figure*}[t]
    \centering
    \subfigure[COVID19 News]{\label{subfig: covid f1}\includegraphics[width=.2\linewidth]{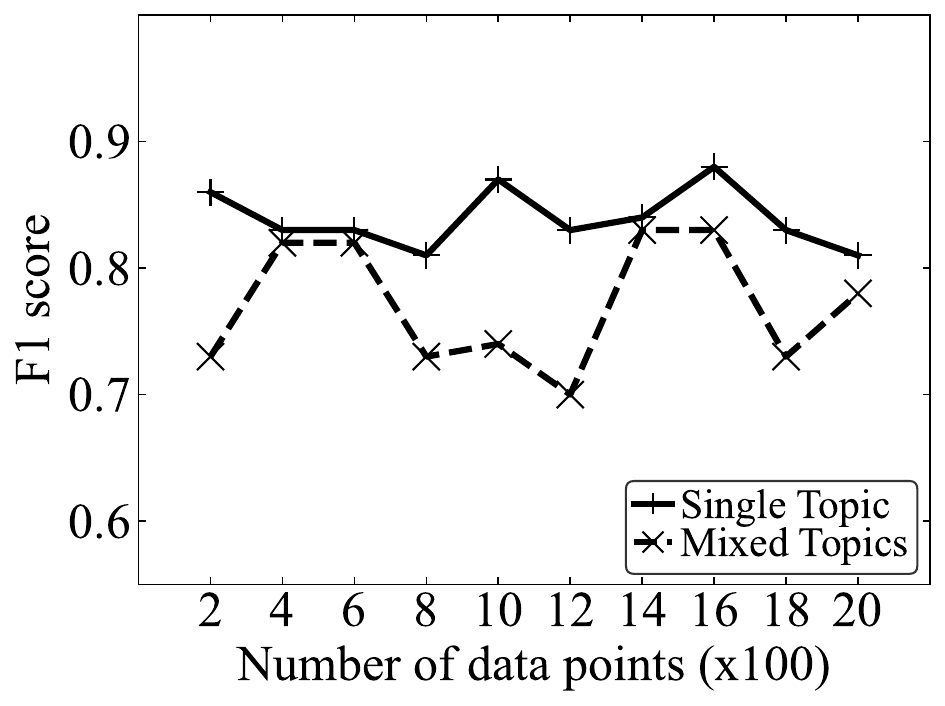}}
    \subfigure[Vaccination News]{\label{subfig: vaccination f1}\includegraphics[width=.2\linewidth]{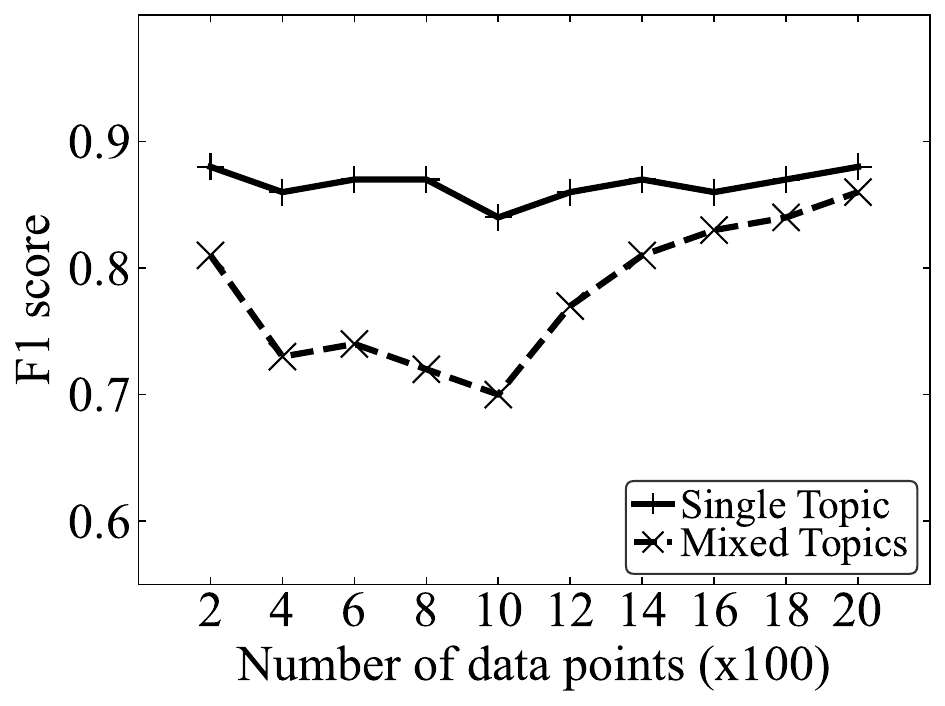}}
    \subfigure[Election News]{\label{subfig: election f1}\includegraphics[width=.2\linewidth]{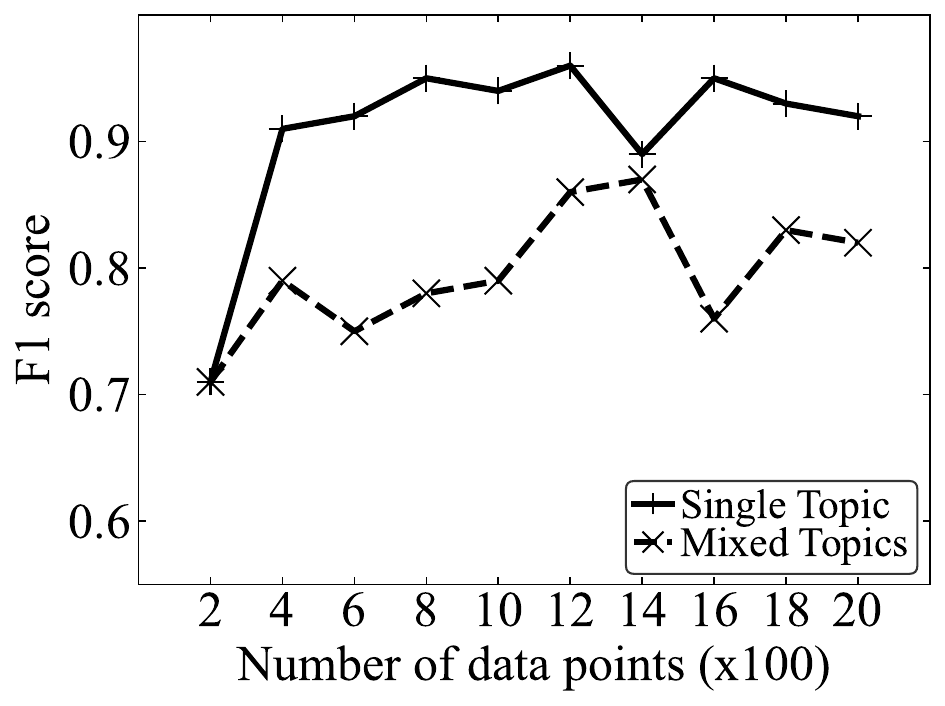}}
    \caption{F1 score on three news datasets: COVID19 news, vaccination news, and election news.}
    \label{fig: three news dataset on f1}
\end{figure*}

\begin{figure*}[t]
    \centering
    \subfigure[Numbers of agents in an aggregation tree.]{\label{subfig: number of nodes}\includegraphics[width=.2\linewidth]{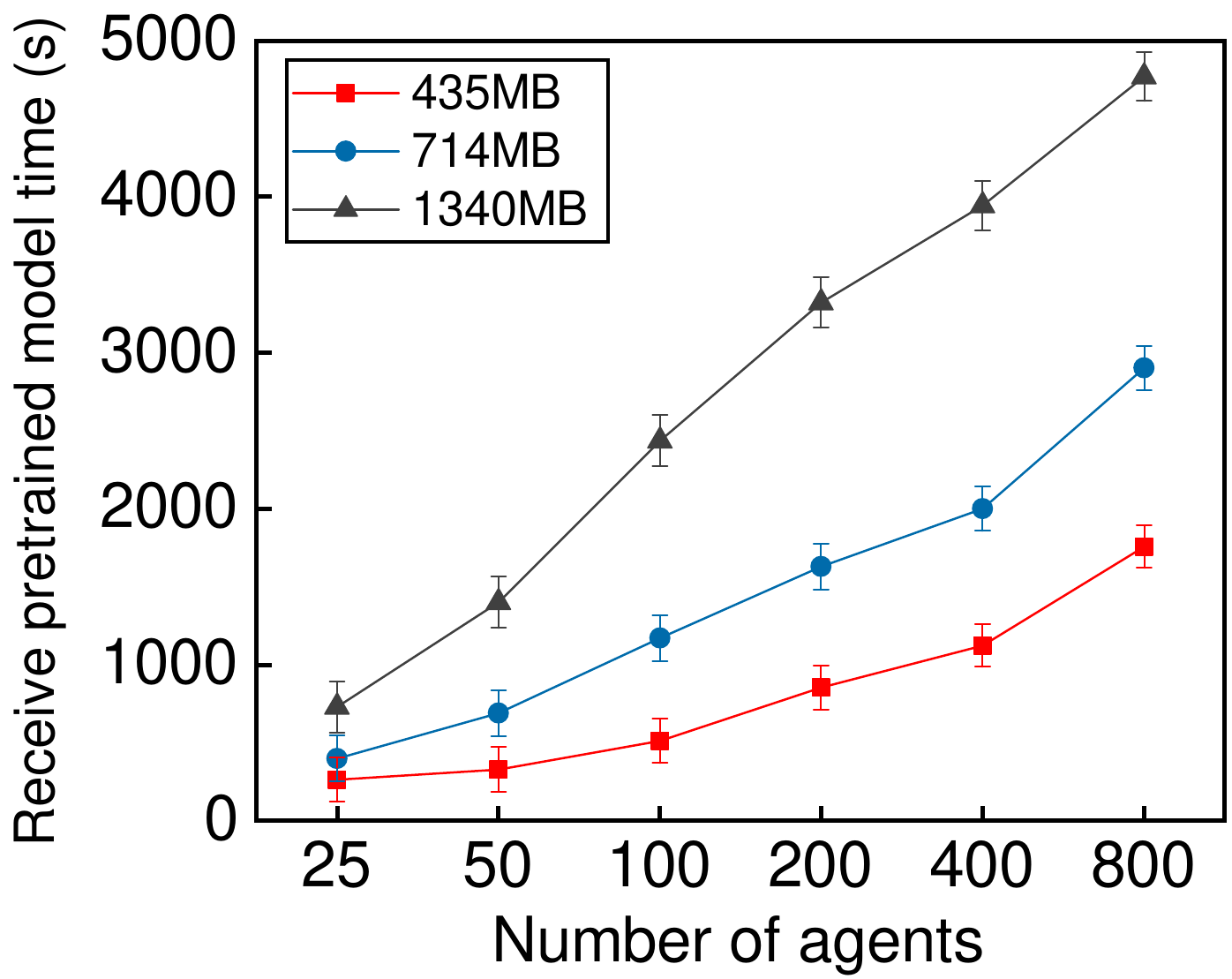}}
    \subfigure[Numbers of aggregation trees for multiple topics.]{\label{subfig: number of trees}\includegraphics[width=.2\linewidth]{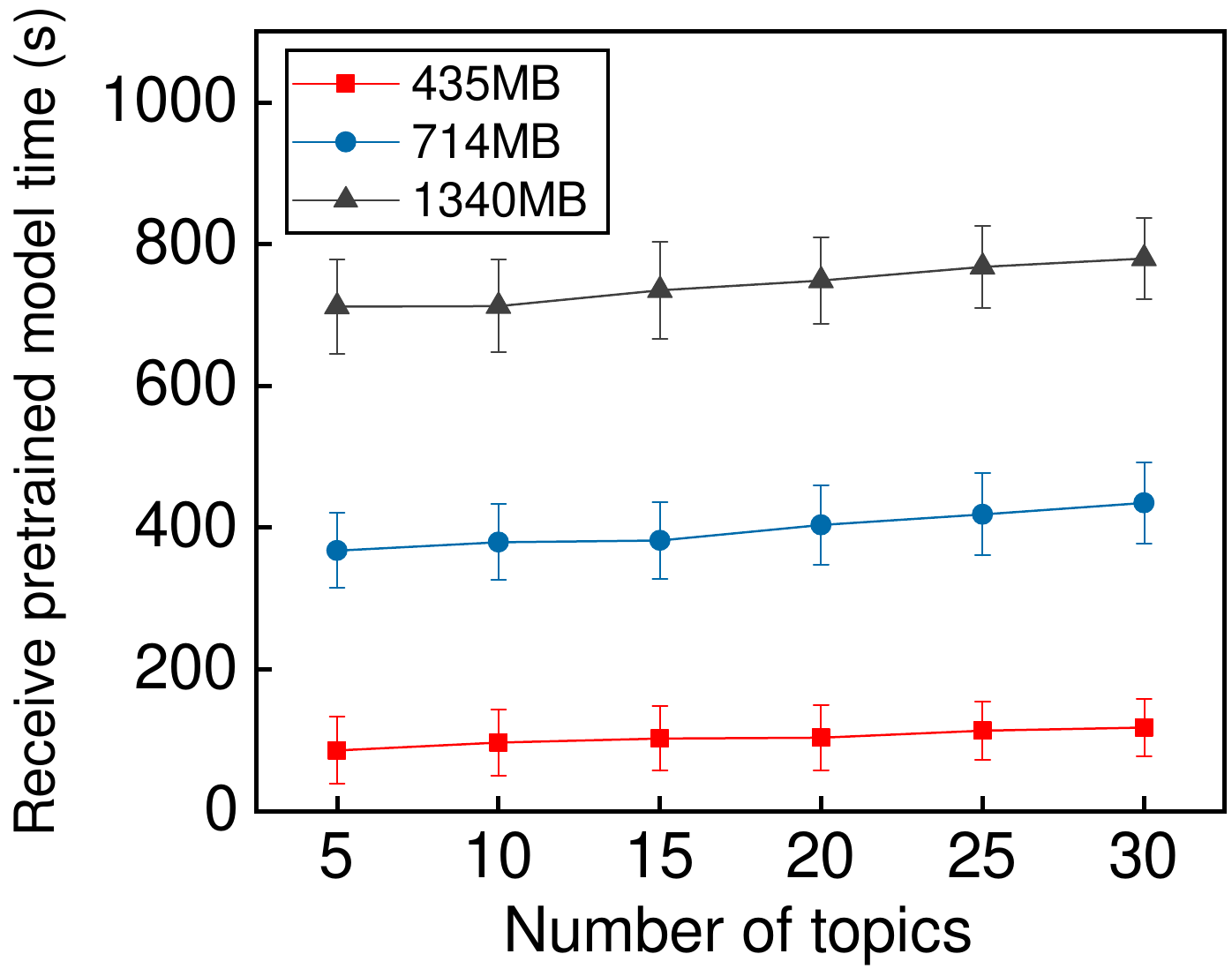}}
    \caption{Received pre-trained model time}
    \label{fig: broadcast latency}
\end{figure*}

Upon receiving the intermediate fine-tuned models from all of the friend leaf nodes, the root $r$ performs final model fine-tuning to yield new global fine-tuned model as follows:
$w^{t+1}_{cla} = w^t_{cla} - \eta \sum_{j\in \mathcal{F}_r} \frac{w^{j,t}_{cla}}{|\mathcal{F}_r|}.$

Compared to {\CSB}, {\DSL} has the following two advantages. One is that the communication and computational bottlenecks can be largely distributed since only a set of leaf nodes are uploading and another set of leaf nodes are computing. Another is that the random noises are unnecessary sine the leaf nodes only share the intermediate fine-tuned models with their friend leaf nodes. Even though the intermediate fine-tuned models will be sent to the leaf nodes two hops away, they cannot know exactly what it is. This is because the intermediate fine-tuned models are first aggregated then sent. It is nearly impossible to the values before addition.

\noindent\textit{Ensemble-based model inference.}
The leaf nodes perform online mode fine-tuning and model inference concurrently. When receiving new data and needing to infer the predictive results for it, the leaf nodes use the fine-tuned model to obtain the predictive results, and then upload the predictive results to the root along the aggregation trees. The root finally aggregates the predictive results in an ensemble-based manner. 

\section{Performance Evaluation} \label{sec: evaluation}


\subsection{Setup}

\paragraph{\textbf{Emulation Deployment.}}
We run all experiments on up to 4 servers, each with 16 Intel Xeon Gold 6130@2.10GHz cores and 256GB of RAM, running GNU/Linux 3.10.0. On top of these servers, we boot up 100 virtual machines to host 1000 DHT nodes in total, each with 4 cores and 8GB of memory, running Linux Ubuntu 16.04.4.

\paragraph{\textbf{Datasets and models.}}
We evaluate {\systemNameAbbr} using three real-world datasets: COVID-19 Fake News Dataset~\cite{covid19_fake_news_dataset}, COVID-19 World Vaccination Progress~\cite{covid19_vaccination_dataset}, and US General Election~\cite{election_dataset}. Each dataset is regarded as an individual topic. 
We train a transformer model (BERT~\cite{kenton2019bert}) to classify the news into true or false classes.


\subsection{Prediction Results}
Figure~\ref{fig: three news dataset on acc} and Figure~\ref{fig: three news dataset on f1} show {\systemNameAbbr}'s prediction accuracy and F1 score, respectively. Single topic means that the training data over the nodes in an aggregation tree belongs to the same dataset, whereas mixed topics mean that each node in an aggregation tree has training data from different datasets. To rule out the effect of the amount of training data on classification accuracy and F1 score, we increase the number of data points in each node from 200 to 2000. We can see that single topic supported by {\systemNameAbbr} achieves higher accuracy and F1 score on three datasets. This is because {\systemNameAbbr} enables each topic to build an individual aggregation tree so that the nodes in different aggregation trees can contribute their model updates to correct global models.

\subsection{Scalability Analysis}

Figure~\ref{subfig: number of nodes} shows the received pre-trained model time for the whole nodes in an aggregation tree to receive different sizes of pre-trained models. Figure~\ref{subfig: number of trees} shows the received pre-trained model time for the whole nodes in multiple aggregation trees to receive different sizes of pre-trained models. We can see that for a single aggregation tree, the received time only increases linearly, not exponentially. This is because the time is limited by the aggregation tree depth $O(\log N)$ by using the DHT-based aggregation tree. On the other hand, we can see that for multiple aggregation trees, the increase of the received time is negligible. This is because  the communication overhead for model dissemination is amortized over the whole nodes in the decentralized overlay.

\section{Conclusion}
{\systemNameAbbr} represents a significant advancement in the field of fake news detection within mobile social networks. By addressing the critical challenges of real-time detection, scalability, and adaptability to mobile network dynamics, {\systemNameAbbr} provides an effective solution to the pervasive issue of fake news in MSNs. The system's innovative use of DHT-based aggregation trees and its dual approach to model fine-tuning enable it to handle the large-scale and rapidly evolving nature of news on these platforms. The ensemble-based model inference and personalized model fine-tuning methods further enhance its efficacy and adaptability to diverse data and detection requirements. Our comprehensive evaluation using three real-world fake news datasets demonstrates {\systemNameAbbr}'s superior performance and functionality, marking it as a pivotal contribution to online fake news detection systems. This system not only defends against the ever-changing landscape of fake news but also sets a precedent for future research and development in this crucial domain.

\begin{acks}
This work is supported by the National Science Foundation under Grants NSF-OAC-23313738, NSF-CAREER-23313737, NSF-SPX-2202859, and NSF-CNS-2322919.
\end{acks}

\bibliographystyle{ACM-Reference-Format}
\bibliography{ref}

\end{document}